\documentclass[fleqn,usenatbib]{mnras}

\usepackage{newtxtext,newtxmath}

\usepackage[T1]{fontenc}

\DeclareRobustCommand{\VAN}[3]{#2}
\let\VANthebibliography\thebibliography
\def\thebibliography{\DeclareRobustCommand{\VAN}[3]{##3}\VANthebibliography}

\usepackage{graphicx}	
\usepackage{amsmath}	
\usepackage{physics}
\usepackage{makecell}

\title[Photoionization feedback in turbulent clouds]{Photoionization feedback in turbulent molecular clouds}

\author[N. S. Sartorio et al.]{
Nina S. Sartorio,$^{1, 2}$\thanks{E-mail: sartorio.nina@gmail.com (NSS)}
Bert Vandenbroucke,$^{3,4}$
Diego Falceta-Goncalves$^{5}$
and Kenneth Wood$^{4}$
\\
$^{1}$Instituto Nacional de Pesquisas Espaciais - Av. dos Astronautas, 1.758 - Jardim da Granja, São José dos Campos - SP, 12227-010, Brazil\\
$^{2}$Institute of Astronomy, University of Cambridge, Madingley Road, Cambridge CB3 0HA, UK\\
$^{3}$ Sterrenkundig Observatorium, Universiteit Gent, Krijgslaan 281, B-9000 Gent, Belgium\\
$^{5}$ Escola de Artes, Ciências e Humanidades, Universidade de São Paulo, Rua Arlindo Bettio 1000, São Paulo, SP, 03828-000, Brazil\\
$^{4}$ SUPA, School of Physics \& Astronomy, University of St Andrews, North Haugh, St Andrews, KY16 9SS, United Kingdom
}

\date{Accepted XXX. Received YYY; in original form ZZZ}

\pubyear{2015}

\begin{document}
\label{firstpage}
\pagerange{\pageref{firstpage}--\pageref{lastpage}}
\maketitle

\begin{abstract}
We present a study of the impact of photoionization feedback from young massive stars on the turbulent statistics of star-forming molecular clouds. This feedback is expected to alter the density structure of molecular clouds and affect future star formation. Using the AMUN-Rad code, we first generate a converged isothermal forced turbulent density structure inside a periodic box. We then insert an ionizing source in this box and inject photoionization energy using a two-temperature pseudo-isothermal equation of state. We study the impact of sources at different locations in the box and of different source luminosities. We find that photoionization has a minor impact on the 2D and 3D statistics of turbulence when turbulence continues to be driven in the presence of a photoionizing source. Photoionization is only able to disrupt the cloud if the turbulence is allowed to decay. In the former scenario, the presence of an HII region inside our model cloud does not lead to a significant impact on observable quantities, independent of the source parameters.
\end{abstract}

\begin{keywords}
turbulence ---radiative transfer --- 
methods: numerical --- ISM: H{\sc ii} regions
\end{keywords}



\section{Introduction} 
\label{sec:intro}

Turbulence is a stochastic flow often associated with a chaotic state of motion due to its non-linear nature and, consequently, hard to model analytically. However, its modelling is key due to ubiquitousness of turbulence flows within both astronomical and non-astronomical contexts. In Astronomy, it has been long known that turbulence is present over a wide range of size scales  \citep{Larson_1981, Solomon_1987}. A turbulent cascade is thought to transfer energies from galactic scales down to scales over ten orders of magnitude smaller \citep{Armstrong_1981}. In particular within molecular clouds, turbulence is of special importance as it is thought to play an essential role in regulating star formation \citep[e.g.][]{MacLow_2004,Ballesteros-Paredes_2007,Federrath_2012,Hennebelle_2013, Orkisz_2017,Mocz_2019,Arzoumanian_2019}. 

In the gravo-turbulent star formation model, turbulence plays two opposing roles depending of the scale being analysed \citep{McKee_2007,Klessen_2011}. On the scales of molecular clouds, turbulent motions could provide support against gravity, preventing a global collapse of the cloud. Contrastingly, at smaller scales, turbulence driven shocks interact to give rise to a hierarchical, fractal-like configuration with larger structures bearing smaller similar structures up to sub-molecular cloud  scales. High density filaments form a net structure with low density voids permeating the space between the denser filaments. Due to their higher densities and relative lower temperatures, within these filaments the forces of gravity may dominate over the internal gas pressure, creating Jeans unstable regions. This makes filaments the ideal place for pre-stellar cores to emerge and star formation to begin. 

The turbulence of the gas within the molecular clouds is well characterised observationally. The velocity dispersion of the gas, obtained by measuring linewidths, indicates that the turbulence is highly supersonic \citep{Zuckerman_1974}. This supersonic nature is quantified by the ratio of the gas velocity in comparison to the speed of sound in that medium, known as the sonic Mach number ($\mathcal{M} = v_{gas}/c_{s}$, where $c_s$ is the sound speed) - which for typical molecular clouds takes values in the range $\mathcal{M} \sim 5-40$ \citep{Zuckerman_1974}.

The value of the sonic Mach number itself is related to the density structure of the molecular cloud. In the simplest case of an isothermal compressible turbulence, the density and velocity across a shock is given by the Rankine-Hugoniot conditions: 

\begin{align}
        \rho_1 v_1 &= \rho_2 v_2\\
        \rho_1 (v_1^2 + c_s^2) &= \rho_2 (v_2^2 + c_s^2).
\end{align}

 Here indices 1 and 2 refer as the conditions on either side of the shock. It is easily derived from Eqs.(1) and (2) that $\rho_2/\rho_1 = \mathcal{M}_1^2$. This implies that the compression factor in an isothermal shock can be arbitrarily high.

Thus we expect the resulting density structure of the turbulence to be related to $\mathcal{M}^2$ \citep{Vazquez-Semadeni_1994, Price_2011, Molina_2012, Burkhart_2015}.

An analytical expression for the density distribution can be reached by considering it as being formed by a succession of multiplicative density perturbations (which are, therefore, additive in the logarithm) created by the turbulence.
Given that the perturbations are independent events, from the central limit theorem, we can expect the logarithm of the density distribution to be a normal distribution \citep{Passot_1998} or equivalently that the mass density probability distribution function (PDF) is log-normal in shape \citep{Vazquez-Semadeni_1994, Padoan_1997}. 

We can characterise this mass distribution by introducing a new variable $s$, defined as the natural logarithm of density divided by the mean density $s = \ln(\rho/\rho_0)$. The mass weighted PDF can then be expressed as:

\begin{equation}
    P(s) = \frac{1}{\sqrt{2\pi}\delta_s} \exp\left(\frac{-(s - \langle s \rangle)^2}{2 \delta_s^2}\right),
\end{equation}

where $\delta_s^2$ is the variance and $``< >"$ represents a mean quantity. This relation yields a fairly good approximation for the mass-density distribution caused by turbulent motions.

Indeed log-normal PDFs have been observed in multiple molecular clouds and used for a number of studies \citep{Kainulainen_2009,Goodman_2009,Hopkins_2013, Burkhart_2015, Mocz_2019}, although in most scenarios they are also accompanied by a tail at high densities which is usually attributed to effect of gravity. Interestingly, this tail seems to be less dominant or even disappear in PDFs for quiescent molecular clouds (which do not have on going star formation)\citep{2013_Schneider}. Simulations have indeed been able to show that, at early times, turbulent motions lead to a density configuration such that the PDF can be precisely represented by a lognormal\citep{2011_Kritsuk}. Nevertheless, once gravity begins to dominate some regions (where the first collapsing objects starts taking place), a power-law tail develops at the high end of the distribution. 

 \citet{Brunt_2015} shows that the emergence of a core and a tail region could also be due to a superposition of the PDFs of the warm diffuse material and of the cold molecular material, instead of being caused by and underlying lognormal PDF with a gravity-induced power-law tail. If we assume that the cold and the warm gas in the ISM can be treated as two separate isothermal fluids, we can expect each to have its own isothermal turbulence distribution. In this case, we would expect that the cold gas has a wider PDF than the warm gas. The superposition of these two PDFs would then appear to be a lognormal PDF with a tail.

By either account, the deviation of the PDF from log-normality is a consequence of the star-formation process either by the effects of gravity or of radiation feedback heating the environment. One way of investigating the latter effect is to simultaneously consider the effects of ionisation and turbulence within a molecular cloud and assess their combined impact on the disturbance of the gas. In this paper we analyse the impact of photoionizing radiation from massive stars in the turbulent structure and probe if ionisation can generate a power-law tail via a suite of numerical simulations.  We enforce an equation of state with adiabatic index $\gamma$ close to 1. In this manner, we able to treat both the cold and the ionized gas roughly as two separate isothermal fluids. We use turbulence-in-a-box type simulations with an isotropically emitting ionizing source to model our molecular cloud and our massive star respectively.

In addition to probing the effect to the density PDF, we can use this study to analyse how other statistics that are usually used to describe turbulence flows change. In \cite{Boneberg_2015}, a molecular cloud with a large number of photoionizing sources was shown to be able to maintain a turbulent state due to photoionizing feedback. In the present work we consider a single point source, which can represent one or more stars depending on the chosen ionizing luminosity. Due to our constant driving of turbulence, we do not attempt to certify if photoionization is capable of generating turbulence, but instead on how we expect the characteristics of the turbulent flow to change when a source is introduced. 

The paper is organised as follows. Section 2 contains the details of our code and numerical simulations, including equations, initial and boundary conditions and parameters used. Section 3 presents the results of statistical analysis of the ISM with and without radiative feedback. Section 4 analyses the results presented in the preceding section and contrasts them with other similar works. We discuss conclusions that can be drawn in the context of molecular cloud integrity and for observations of interstellar turbulence. In this section we also scrutinise the impact of our simplified turbulence model, speculating on the possible differences that may arise if we were to consider a model that includes physical aspects ignored here, as self-gravity and MHD turbulence. We conclude with a summary of the conclusions in Section 5.

\section{Method}

The simulations presented in this paper are performed with the AMUN code  \citep{Kowal_2011, Kowal_2017} coupled to a radiative transfer code similar to \citet{Wood_2004, Vandenbroucke_2018}. The code uses a  Godunov-type scheme with an HLL Riemann solver and second order Runge-Kutta time integration methods to solve the hydrodynamic equations:

\begin{equation}
    \begin{aligned}
    \frac{\partial \rho}{\partial t} & + \grad \cdot{\rho \textbf{v}} = 0 \\
    \frac{\partial \rho \textbf{v}}{\partial t} & + \grad \cdot\left[\rho\textbf{v}\textbf{v} + p\right] = f
    \end{aligned}
\end{equation}

where $\rho$, $v$, $p$ are respectively density, velocity and pressure of the gas and $t$ is the time. Here the source term $f$ is responsible for the spectral velocity forcing of the turbulence which is done in a similar manner to the one proposed by \citet{Alvelius_1999}. This method uses a stochastic force to numerically induce a state of homogeneous turbulence for which the statistics is invariant over time. The forcing is applied in Fourier space where it is concentrated at small wave numbers/large scales ( for our simulations the injection scale is chosen to be half of the box size corresponding to 5pc ). The forcing components have random phases at each step and the correlation between them and velocity is removed, thereby ensuring there is no correlation between the velocity field and the forcing at any temporal or spatial scale. The result is a forced compressible and purely solenoidal turbulence.

This forcing is introduced to a cubic box with initially uniform density of $1000$ particles $\rm{cm}^{-3}$ with resolution $256^3$ and periodic boundary conditions. We apply a time-step constraint based on the Courant stability criterion and we have a global timestep for our simulation box. It should be noted that although we use a barotropic equation of state with $p\propto \rho^\gamma$, we set the adiabatic index $\gamma = 1.0001$ effectively mimicking an isothermal scenario.

For simplicity, the material in the simulated molecular cloud in our radiation-hydrodynamics runs is assumed to be composed solely of atomic hydrogen, with no heavier elements or dust. As our aim is to determine the dynamical impact of photoionization alone, assuming pure hydrogen is adequate. The low abundances of heavier elements make their opacities much lower than that of hydrogen  and, consequently, they do not significantly influence the radiation transport and the resulting ionization structure of hydrogen. It is worth noting that, in the colder denser regions in the filaments, hydrogen will be often found in molecular form. One would expect that 13.6 eV photons would be able to traverse much larger distances in molecular gas than in the atomic gas modelled here, since they have a small probability of interacting wiht $H_2$. However this effect is substantially diminished due the dissociation of H2 by other UV radiation on $H_2$ dissociation (eg. Solomon process and $H_2$ photoionization followed by dissociative recombination) \citep[see for instance][]{Baczynski_2015}. Ignoring $H_2$ dissociation allows the majority of 13.6 eV photons to completely escape our molecular clouds without interacting, since the atomic hydrogen abundance would be lower than in reality. Thus, while in practice treating the gas as being atomic rather that molecular is likely to underestimate the reach of re-emitted radiation, this should not strongly affect the size of our ionized region. A proper treatment of this effect would require an appropriate tracing of formation and destruction of molecular hydrogen and include the other relevant photon energies, which is beyond the scope of this paper. 
The one place where elements other than hydrogen play a pivotal role is in the determination of the temperature of the gas. Including metals is vital for estimating cooling rates and, thus, with their absence, we are not able to self-consistently determine the temperature. Instead, we use a two temperature approximation in which fully neutral cells always have a set temperature of $T_n =$ 100K, whereas fully ionized cells have a fixed temperature of $T_i = 10^4$ K.
Partially ionized cells have a temperature equivalent to a linear interpolation between the fully ionized and the neutral state: 

\begin{equation}
    T_{cell} = \frac{T_{i}n_H^+ + T_{n}n_H^0 }{n_H^+ + n_H^0}
\end{equation}

where $n_H^0$ and $n_H^+$ are the number of hydrogen atoms in the neutral and on the ionized state for that cell respectively.
This method is shown to reproduce well the temperature profile for H{\sc ii} regions \citep{Vandenbroucke_2020}.
The two temperature approximation also has the advantage that it is computationally less demanding. The temperatures $T_n$ and $T_i$ are based on the results we get by using a more complex version of the code which does perform the cooling self-consistently \citep{Wood_2004}. While $10^4$K is a good approximation for the value of ionized gas, the neutral temperature can be as low as $10$K in some molecular cloud regions, which makes our choice of $100$K for neutral gas appear to be excessively high.   However, it is worth pointing out that the filling factor of gas at temperatures of 10-20K is small compared to our cell sizes \citep{Kulkarni_1988}. In this sense 100K reflects the average temperature of each cell better. Furthermore, the choice of $T_n$ should not have a large impact in our simulations as it affects mostly the gravitational stability of the gas (colder gas would fragment more easily) and these simulations do not trace fragmentation or possess self-gravity. While this may imply that our turbulent velocities are higher than the ones found in the ISM (since we would be over-estimating $c_s$), because we set the Mach number of our simulation to $\mathcal{M} = 6$ the compression of the gas and resulting density structures would be the same as if had ran then at a lower average temperature. Thus, though the temperature choices may alter slightly the quantitative value of the statistics, the qualitative impact of radiative feedback can still be probed.

The photoionization is performed using a Monte Carlo radiation transfer code which employs a model whereby ionizing photon-packets are isotropically emitted from a source. As we do not emit photon packets with distinct energies, we do not sample the stellar spectrum. Therefore, in order to correctly account for the wide range of frequencies of the stellar radiation, we weigh the cross section based on the flux at each frequency:

\begin{equation*}
    <\sigma > = \int_{f_{min}}^{f_{max}}\frac{\sigma(\nu)B_\nu}{h\nu} d\nu \bigg{/} \int_{f_{min}}^{f_{max}}\frac{B_\nu}{h\nu}d\nu
\end{equation*}

with the minimum ionizing energy $f_{min} = 13.6$ eV and $f_{max} =124$eV (the largest energy for UV radiation), $\sigma(\nu)$ is the cross section at a frequency $\nu$ and $B_\nu$ is the Plank function:

\begin{equation*}
    B_\nu(T) =\frac{2h\nu^3}{c^2(e^{h\nu/kT}-1)},
\end{equation*}

The flux is assumed to be that of a black body with a temperature of $30,000$K corresponding to an O-type star. This results in a flux averaged cross section $<\sigma> =0.44 \sigma_0$, where  $\sigma_0$ is the cross section for the $1^2S$ level of a neutral hydrogen atom ($\sigma_0 = 6.30\times 10^{-18}~{\rm{cm}}^2$).

The ionizing packets are propagated through a discrete density grid until a randomly sampled optical depth is reached. At this point two things can happen to the photon: either it is absorbed (thereby terminating the packet), or it is re-emitted as another ionizing photon with probability $\alpha_1/\alpha_A \approx 0.36$. Here  $\alpha_1$ is the recombination rate to ground level and $\alpha_A $ to all levels for a gas at 10,000K, which is the characteristic temperature of H{\sc ii} regions (such that $\alpha_A \approx 4.9 \times 10^{-13} \rm{cm}^3/\rm{s}$ and  $\alpha_B \approx 3.1 \times 10^{-13} \rm{cm}^3/\rm{s}$; \citealp{Osterbrock_2006}).

We do not adopt the so called on the spot approximation, whereby re-emitted photons are absorbed locally. Instead we treat re-emitted ionizing photons in the same way as source photons, thereby simulating a diffuse field. The only difference for diffuse photons is that they will have a set frequency of 13.6~eV, meaning that the cross section for these photons is not the flux averaged one, but simply $\sigma_0$ \citep{Wood_2000}. Photon packets that escape the simulation box are removed from the system. For each cell that gets traversed by a photon we add to a cell dependant intensity counter an amount proportional to the size of the path the photon packet travelled within the cell. After all the packets have been emitted and propagated (in these simulations we use $10^6$ photon packets), the mean intensity within each cell is used to compute how much of gas within the cell was ionized \citep{Lucy_1999}. The opacity grid is then updated to take into account that the ionized gas is transparent to the incoming radiation. We repeat the emission and propagation of photons ten times to get a converged ionized result for each Monte Carlo time-step.

The simulation runs are divided in two stages. In the first stage, we run a single simulation (without radiation) of a turbulent box with a Mach number of 6. At this point the simulation is scale free, but for this paper it is used as a proxy for a region of 10pc$^3$, which corresponds to a simulated cloud mass of 1000 $M_\odot$. This hydrodynamical simulation is run until a statistically stationary state for the turbulence is achieved which is checked by certifying that the dissipation rate is equal to the power input from the forcing. The final snapshot of this first hydrodynamical simulation is then used as the basis for the second stage, which consists of 5 distinct simulations. One of the follow up simulations is a control run in which no source is inserted and turbulence just continues being forced in a hydrodynamical box. 
We refer to this simulation as the ``HD run". In the other four simulations, an isotropically emitting ionizing source is added. We continue driving the turbulence, but now we perform radiative transfer in every hydrodynamical timestep. These are referred as the ``RHD runs". For all simulations the boundary conditions adopted for the hydrodynamics are periodic, while the radiation is treated with open boundary conditions.
Each RHD simulation differs in regard to the position of the source (being a pointlike source in either a high or a low density region) and the luminosity (varying from 1 or 10 times the usual ionizing luminosity of an O star, that is, $10^{49} \rm{s}^{-1}$. The sources themselves remain stationary throughout the simulations, but the gas around them, including the filaments, are allowed to move freely.

\section{Results}

In this section we present and analyse statistical differences between our runs. All runs start form a turbulent box with stationary statistical properties in which we insert an ionizing source in a region of interest and then evolve the system with full radiative hydrodynamics.

The runs vary in two aspects: the luminosity of the source and the position of the source. The ionizing luminosity is set to either $Q =10^{49}\sc{s}^{-1}$ (equivalent to value expected for a single O star) or one order of magnitude larger, $Q =10^{50}\sc{s}^{-1}$, which can be thought of as a small cluster. The source is put in either a filament or in a low density cavity. The former is chosen by selecting the densest cell in the central slice of the simulation box and then verifying that this cell's closest neighbours also have a high density. 
This is aimed to place the star initially in an environment similar to the one in which it will be formed, as we lack the resolution and physics in the present simulations to stipulate the necessary criteria to simulate star formation consistently.  The sources in a low density cavity, were placed such that they were approximately at the center of the cavity, as to represent the extreme opposite of the source placed in a filament.

We name our runs according to the luminosity and placement of the source as laid out in Table \ref{tab:runs}.

\begin{table*}
 \begin{tabular}{lllll}
  \hline
  Name & Q ($10^{49}$ photons/s)& Positioning of the source & \makecell{Average dens. at \\source position $\rho/\rho_0$} & $R_S$ (pc)\\
  \hline
  HDL1 & 1 & filament  & 9.31 & 0.113\\
 HDL10 & 10 & filament& 9.31 & 0.243\\
 LDL1 & 1 & void & 3.35$\times10^{-2}$ & 5.65\\
 LDL10 & 10 & void& 3.35$\times10^{-2}$ & 12.19\\
  \hline
 \end{tabular}
 \caption{ Parameters used and expected Stromgren sphere radii for each one of our simulated cases. This computation assumes an ionized region forming in a uniform density region with number density equal to the average one (1000 cells) around the position of the source}
  \label{tab:runs}
\end{table*}

\begin{figure*}
    \centering
    \includegraphics[width=2\columnwidth]{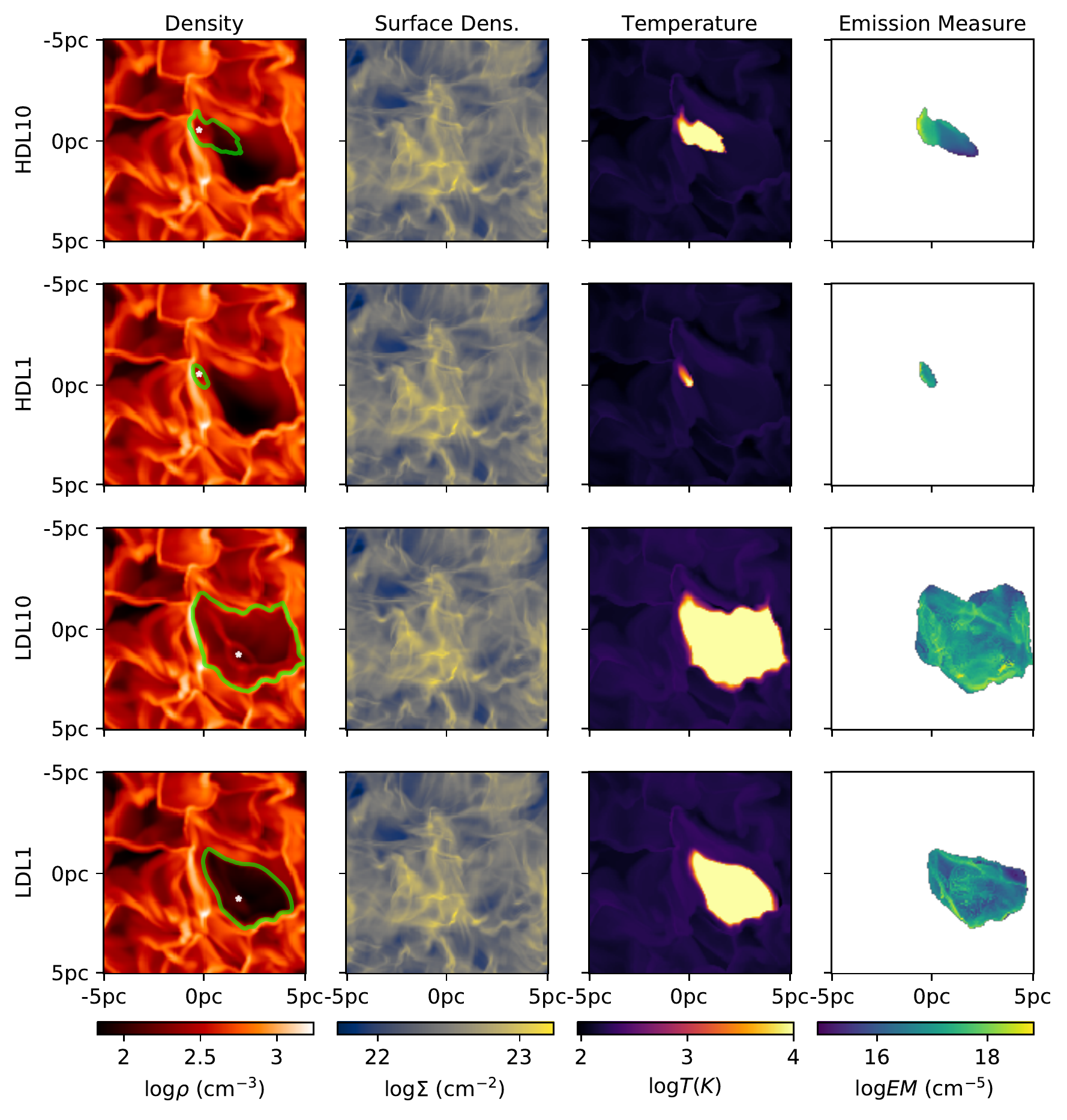}
    \caption{Columns (from left to right): density slice, surface density, temperature slice and emission measure. Slices are taken at a distance of 0.25 pc from the source perpendicular to the slice at a time 0.8 Myr from the introduction of the source. The green contour in the first column shows the boundary of the H{\sc ii} region where material is 80\% ionized.  Rows (from top to bottom): Star placed in a filament with $Q = 10^{50}\rm{s}^{-1}$, star placed in a filament with $Q = 10^{49}\rm{s}^{-1}$, star in a void $Q = 10^{50}\rm{s}^{-1}$ and star in a void $Q = 10^{49}\rm{s}^{-1}$.  The number of cells ionized is 31,272 ($0.19\%$ of box), 1024 ($0.006\%$ of box), 1,275,984 ($7.6\%$ of box) and 538,864 cells ($3.2\%$ of box) for HDL10, HDL1, LDL10 and LDL1 respectively. }
    \label{fig:slice_comparison}
\end{figure*}

Figure \ref{fig:slice_comparison} illustrates the qualitative differences between the runs after 0.8 Myrs from the introduction of the source to the turbulent grid. The top two rows refer to the runs in which the source is placed in a filament while the bottom rows refer to runs with a source in a void.

In order to understand the results is useful to first consider what happens when an isotropic source of ionizing radiation is placed in a uniform medium. In this scenario the source will almost instantly ionize a sphere of material, known as the Stromgren sphere. The size of this sphere is dictated by the balance between the recombination rate of ionized atoms to a neutral and the flux of ionizing photons. Its radius can, thus, be derived from the ionization balance equation to be:
\begin{equation}
    R_s = \frac{3}{4\pi}\frac{Q}{n^2\alpha_B}
\end{equation}
where $Q$ is the ionizing luminosity in photons/$\rm{s}^{-1}$, $n$ is the number density of electrons (which for an originally neutral gas will be the same as the number density of both the initial neutral gas and the ionized atoms within the Stromgren sphere) and $\alpha_B$ is the recombination rate to the ground state which for gas at 10,000 K (our assumed ionized temperature) is $\alpha_B = 1.3\times10^{-13} cm^{3}/s$.
After this sphere forms this bubble expands due to the large internal pressure of the hot ionized gas in comparison to its cold neutral surroundings. This expansion leads to the formation of a shock front in which gas accumulates and gets denser. This behaviour is well studied and is supported by a number of analytical as well as numerical works \citep{Spitzer_1978,Hosokawa_2006,Raga_2012, Bisbas_2015, Williams_2018}.

In Table \ref{tab:runs} we show the average density of a cube of $10^3$ cells centered at the source position for each run. We show, for these densities and the ionizing luminosities adopted for the individual runs the initial Stromgren sphere that would form. It can be seen that the final H{\sc ii} shown in Figure \ref{fig:slice_comparison} is smaller than these predicted spheres for LDL1 and LDL10. As a result, in our simulations, the cavity gets quickly ionized and the H{\sc ii} region is delimited by the dense filaments that it finds standing in the way, hindering its expansion. Consequently, instead of giving rise to a shock front, the ionization pressure is simply able to push the filaments slightly. Wherever pressure causes the compression of gas to higher densities there is no particular density that gets affected because the ionization front is surrounded by gas at a large range of densities.

In both HDL1 and HDL10 simulations the expansion of the ionized region is promptly hampered by the densest regions in the filament. In the case where the ionizing emission is consistent with the one of a single massive star (HDL1) the H{\sc ii} region finds itself trapped within the filamentary region, while in the case where the ionizing luminosity is that of a few massive stars (HDL10), the filament is dismantled to one side, giving the opportunity for the H{\sc ii} region to expand into a low density region. Note, however, that the filament itself is not completely destroyed.

In the case for a single star placed in a void (LDL1), all the low density gas of the cavity gets ionized, but the source itself is not strong enough to displace any of the filaments from their initial position. In the increased luminosity ( $Q=10^{50}\rm{s}^{-1}$) case, the pressure from the ionized region is large enough to push into the lower density filaments, although the higher density filament (at the image centre) remains virtually unaltered. In the higher ionizing luminosity run, we can also see that some gas has been pushed and compressed in the direction perpendicular to the slice, which can be seen by the fact the cavity now appears to have a higher density. The slices in Figure \ref{fig:slice_comparison} were taken at a small offset in $z$ of 0.25~pc with respect to the source position to illustrate this behaviour.

An interesting result arises when trying to compare the surface density of the different runs: for all runs it looks effectively identical, a point we will later support by the 2D statistics for the runs later in this section. The ionized regions have fairly uniform temperature, which sharply decays at the H{\sc ii} region boundary.

In the final column of Figure \ref{fig:slice_comparison}, we plot the emission measure, defined as:

\begin{equation}
   EM = \int n_e^2 dz
\end{equation}

where $n_e$ is the electron density and $z$ is the direction perpendicular to the line of sight. This is proportional to the $H_\alpha$ surface brightness, providing a notion of how the ionized region would appear in an observation. 
The EM matches the shape of the ionized region in the slices relatively well, in particular for the models where the ionizing source is in a filament. This is expected as the slices are taken close to the plane in which the ionized region is at its widest and, thus, the shape integrated emission along the LOS will have a similar shape.
The shape is more distinct for the models with an ionizing source in a low density region. In addition the EM shows filamentary structures similar to those in the surface density plot, indicating that at least some intermediate density filaments are (partially) ionized.

\subsection{The 3D density PDF}

As discussed in earlier sections, the high end tail of the PDF of the  density of the simulated molecular cloud can provide hints on the gas that may be prone to form stars. Here we consider if the tail of the density PDF could arise by the existence of an ionized region such that the full density PDF is a combination of two isothermal turbulence PDFs: one for the neutral and one for the ionized gas. 

\subsubsection{Comparison between ionized and neutral gas PDFs}
\label{subsec:PDFdens1}

\begin{figure*}
  \includegraphics[width=2\columnwidth]{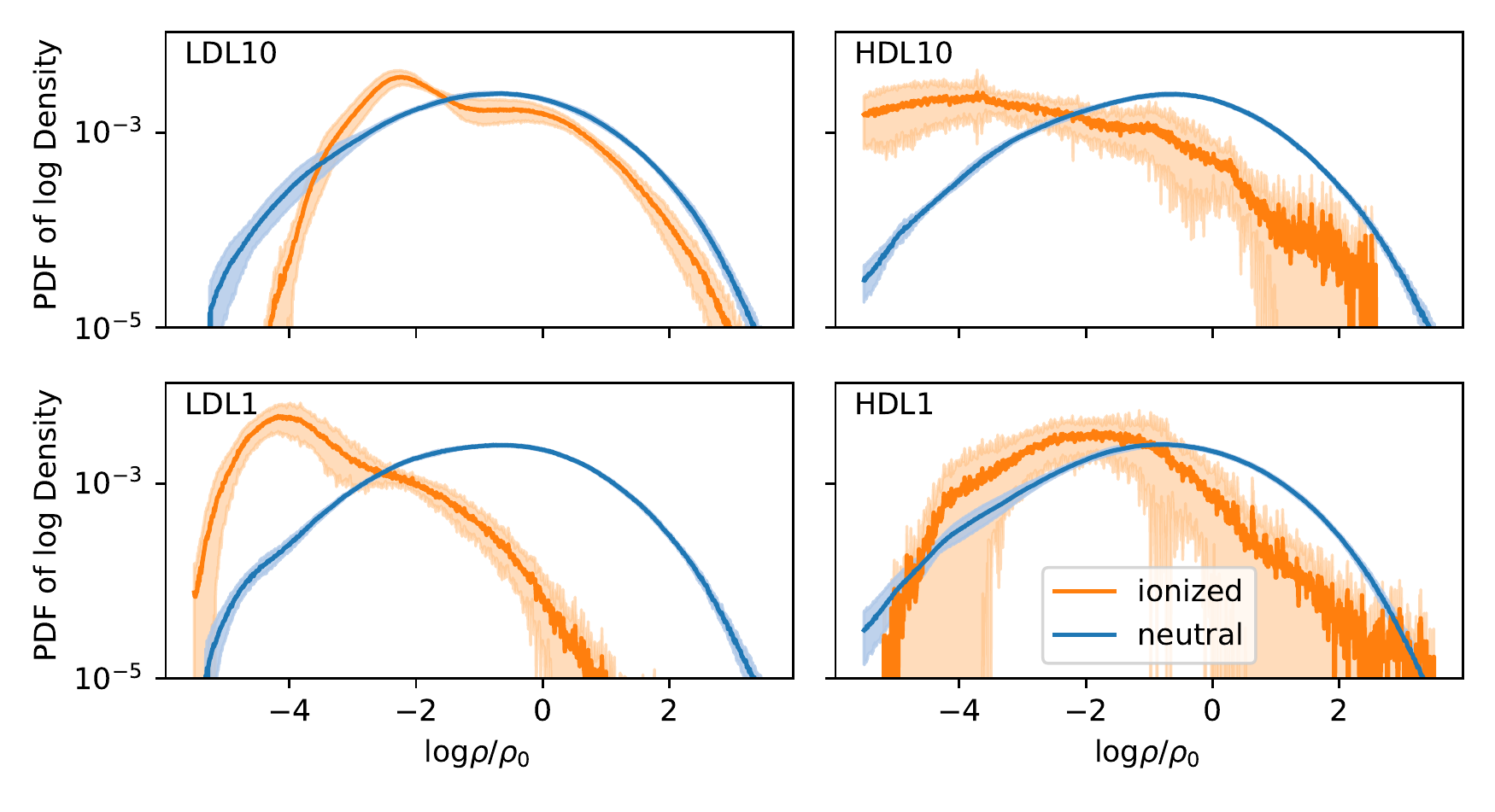}  
\caption{Normalized average density PDF of our four simulations; two where the ionizing source is placed in a low density region (LDL10 and LDL1) and two where the ionizing source is placed in a high density region (HDL10 and HDL1). The averages are taken 
through the first 50 snapshots (each at 0.1 Myr apart), from the introduction of the source. The shaded regions show the spread and the solid lines show the average value for each density. The top plots correspond to simulations with the ionizing luminosity set to $Q = 10^{50} \rm{s}^{-1}$ photons, whereas the bottom plots show the result when ionizing luminosity is decreased by an order of magnitude ($Q = 10^{49} \rm{s}^{-1}$). The blue lines show the PDF obtained by only considering the neutral gas, whereas the orange lines consider only the ionized gas within the same set of simulations.}
\label{fig:PDF_3d_2}
\end{figure*}

In Figure \ref{fig:PDF_3d_2} we plot a the PDF of the ionized and the PDF neutral gas within the same simulation. These PDFs were averaged over  50 snapshots 0.1 Myrs apart from the introduction of the source and then normalized. This averaging is used for all PDFs and CDFs presented unless stated otherwise.
The two plots on the left correspond to the simulations with a source at a void: LDL1 (bottom) and LDL10 (top). For LDL1, the source ionizes mainly very low density gas, having no impact in the high end tail of the PDF.  In the LDL10 simulation, a lot of denser gas is ionized. However, the gas that was compressed to higher densities by the H{\sc ii} region only reaches number densities of $0.1 \rm{cm}^{-3}$, as can be seen by where the PDF of the ionized gas peaks. The high density end of the ionized region follows the same profile as the neutral one, as the source only ionized gas that was already dense but exerted no considerable extra compression.

We can compare these with the plots on the right of Figure \ref{fig:PDF_3d_2}, which shows equivalent plots for the sources placed in a high density region (filament). The first noticeable difference is that the curve for the ionized gas is a lot less smooth. This owes to the fact that these ionized region are a lot smaller (as can be seen from Figure \ref{fig:slice_comparison}), which reflects as having a lesser number of cells at each density, making the PDF appear more noisy. The bottom plot represents run HDL1 in which the ionized region does not manage to perforate the filament and, therefore, remains restrained on all sides by the filament. In this case, most of the ionized material finds itself in intermediate densities which corresponds to the density of the cavity carved by the ionizing radiation coming from the source. Some of the highest density cells also manage to get ionized at the edge of the H{\sc ii} region. 
In the upper panel, we depict the case for HDL10. As it can be seen, increasing the luminosity has two effects: (i) a lot more low density material is ionized and (ii) the densest  ionized cells are no longer as dense as the ones in the neutral PDF. The former is easily accounted for as, once the filament is disrupted, a pathway to a low density cavity opens and suddenly a lot of lower density material is exposed to ionizing radiation.  As recombination rates in this gas are low, this region is readily ionized. The lack of highly ionized cells can be understood by looking back at Fig. \ref{fig:slice_comparison}: in the highest luminosity case, the ionizing source ionizes the full width of the filament on the top left corner, where the densest cells lie. Once the filament is ionized, there is no longer a density gradient opposing the pressure of the ionized region. As a result, the ionized dense filament inflates due to the higher temperature of the gas, relaxing to lower densities.

\subsubsection{Comparison between runs with and without an ionizing source}
\label{subsec:PDFdens2}

\begin{figure*}
    \centering
    \includegraphics[width = 1\columnwidth]{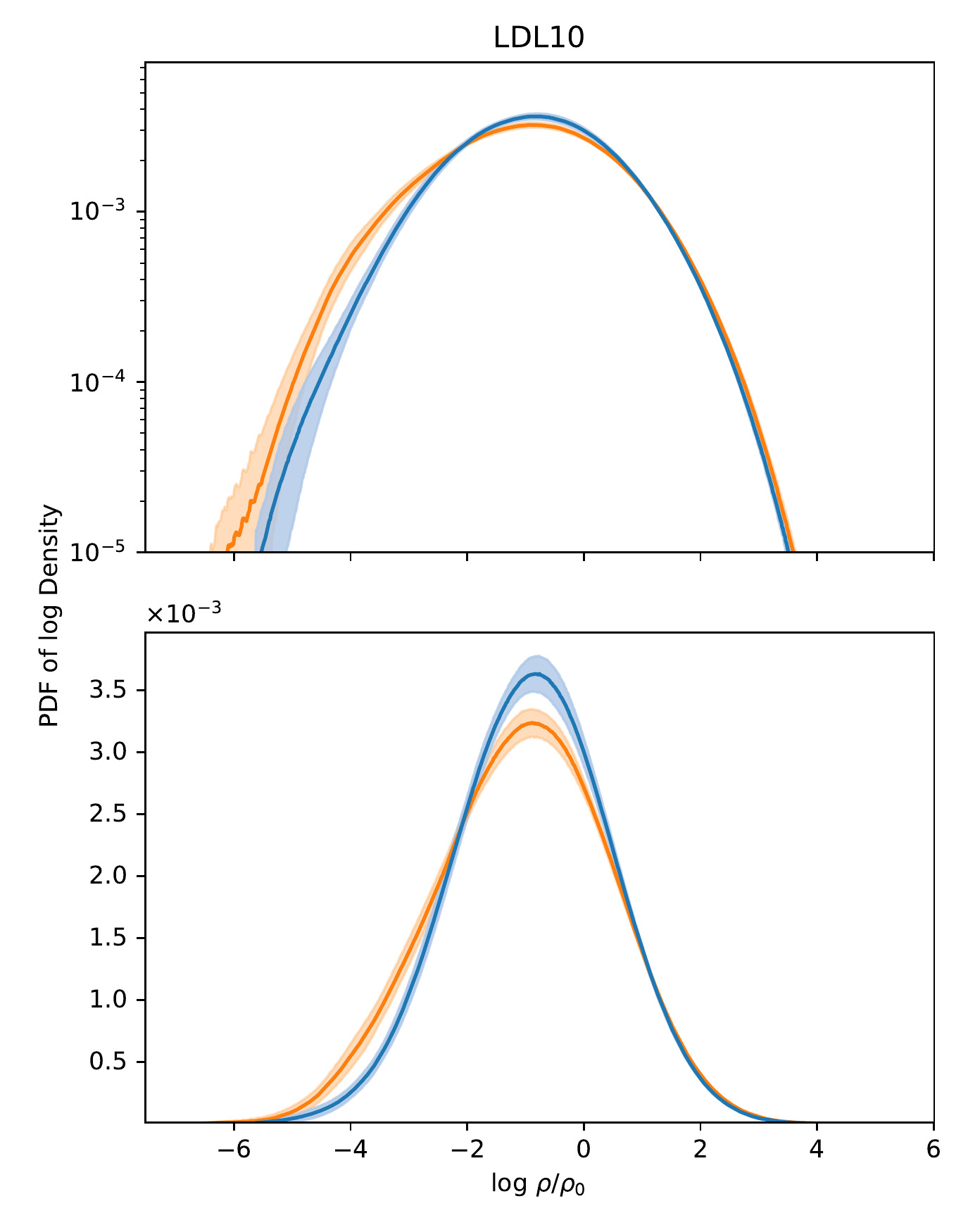}
    \includegraphics[width=0.95\columnwidth]{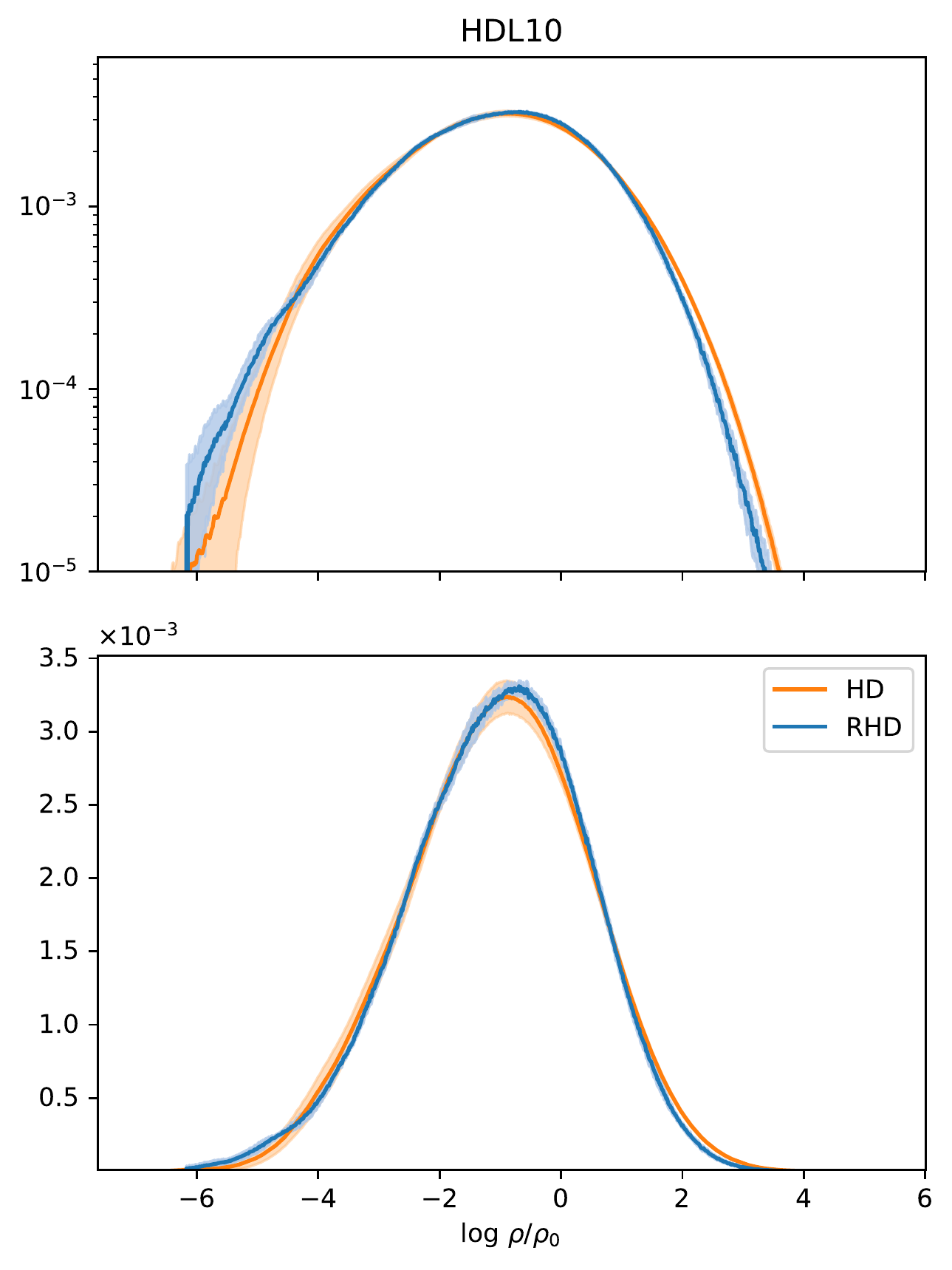}
    \caption{Normalized density PDFs for simulation LDL10, which has the highest dynamical impact. Top figure has a logarithmic scaling as to highlight the behaviour at low and high density tails, while the bottom plot has a linear scaling which shows more clearly the effect around the peak region.}
    \label{fig:density_pdf_snapshot_comparison_LD}
\end{figure*}

Even though the difference between the PDF of the ionized and the neutral gas is clear in every run, the difference between the PDF of the control (HD) run and the PDF for the total density (both neutral and ionized gas) of the runs with ionization is much smaller. This is due to the fact that in most scenarios the ionized region does not compress or dilute a large fraction of the gas in our simulation box, having thus a limited impact on the overall shape of the final PDF. In Figure \ref{fig:density_pdf_snapshot_comparison_LD} we show the average PDF for all 50 snapshots for the run in which no source was inserted (HD run) and the average PDF for the RHD runs LDL10 and HDL10. 

In LDL10 we can see that the high end tail of the PDF is the same in both the RHD and the HD runs.  This indicates that, even if some dense gas is ionized, the number of cells containing the highest density values has not changed. The largest changes are seen in the low density tail, where the HD run has more cells, and at the intermediate densities which have larger values for the ionized run. This shows clearly that, in the high luminosity scenario, some of the low density gas in the cavity can be compressed to higher densities by the incoming ionizing flux. In the case at hand, however, no gas is compressed to densities so high that we would expect it to potentialy be gravitationally unstable. As such, the photoionization feedback in the cases depicted here should not influence star formation rates. This does not, however, rule out the possibility that the interaction of the material being pushed by multiple H{\sc ii} regions could create (star forming) high density sites. It should be observed that, as seen in \ref{subsec:PDFdens1}, part of the high density tail of the PDF can be ionized while remaining dense. In turn, this implies that care must be taken when connecting the high density tail of the density PDF to star formation, as high density warm ionized gas is not a candidate to form new stars.

In the comparison between HDL10 and the control run, the scenario is inverted at the low density wing: the HD run has less low density cells than the RHD run. In addition, in this case there is a marginal difference in the high density end of the PDF, with the RHD run having less cells with the highest density values. Here the increase in the low density cells arises because the ionizing source opens a low density cavity in the middle of the filament that did not previously exist. Simultaneously, in this RHD run the densest region of the filament is ionized by the source. Due to its higher gas temperature, the filament expands due to gas pressure, becoming less dense in the process.
It should be noted that, although this effect would happen in any simulation, the lack of self gravity in this work may intensify the filament rarefaction as there is no force counterbalancing the thermal pressure. This accounts for the lack of dense cells in the right wing of the RHD PDF and the increase of cells with densities at its upper middle region.

\subsection{Surface density PDF}

\begin{figure}
    \centering
    \includegraphics[width=\columnwidth]{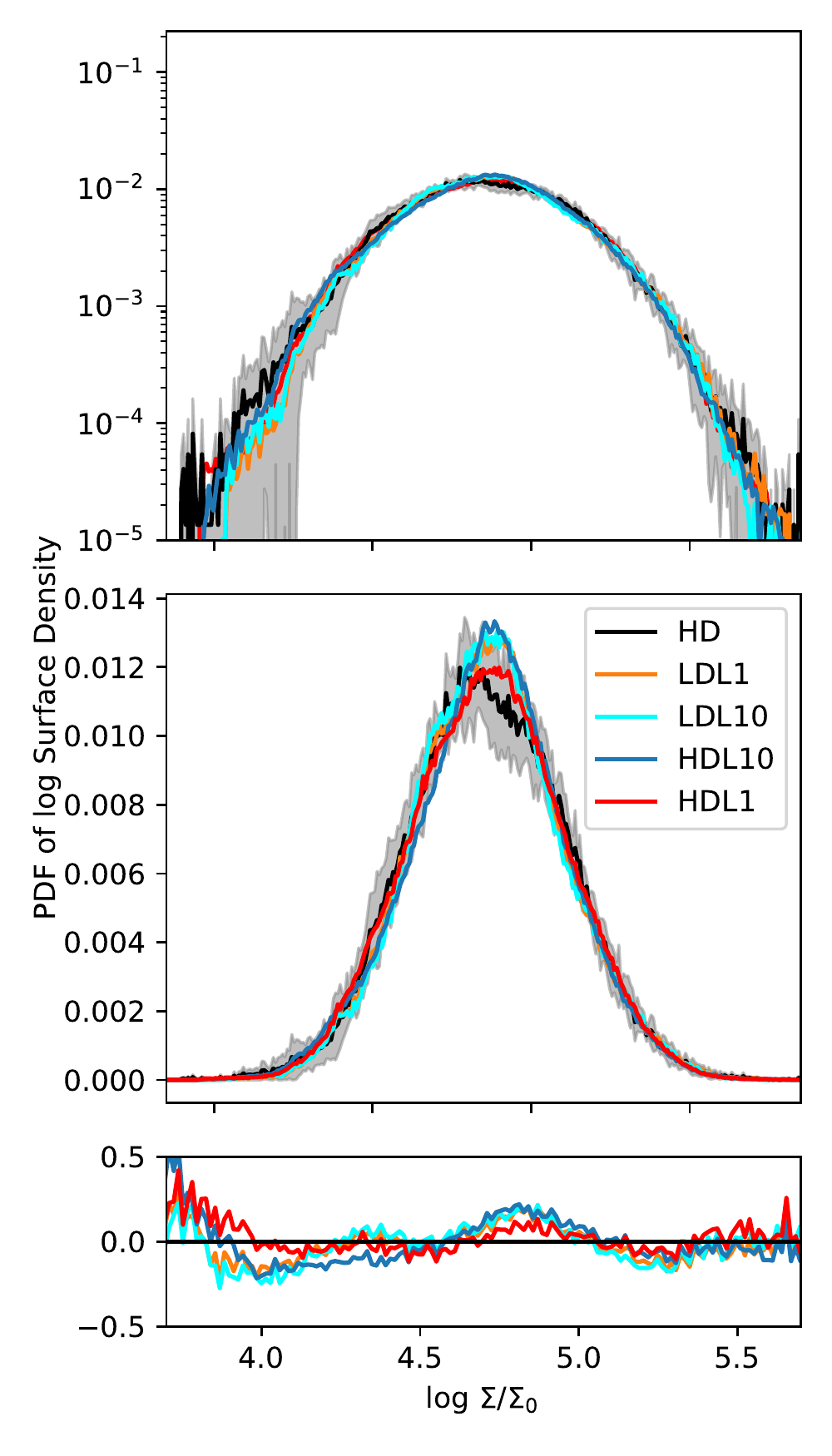}
    \caption{The top two plots show the average surface density PDF for each of the RHD runs (coloured lines) in comparison with the spread for the HD run (in grey) in a logarithmic and linear scaling respectively. The bottom plot shows the fractional change of the average PDF for each RHD run with respect to the control simulation, such that a value of zero indicates no difference between the RHD and the HD run. }
    \label{fig: PDF_surf_dens}
\end{figure}

Although the 3D statistics show us what is actually happening to the gas, we are often constrained to a 2D picture of molecular clouds as observations produce column density maps along the observer's line of sight (LOS). 
In Figure \ref{fig: PDF_surf_dens} we show the surface density obtained for each of the  RHD runs in comparison with the spread for the HD run. The difference is mostly accentuated at average densities, while at the edges the RHD curves seem to be well within the spread for the HD run. The deviation of the RHD values from the average HD value is in general small, and owes to low volume in which densities are changed due to the interaction with H{\sc ii} region.

\subsection{Velocity Power Spectrum}

The power spectrum is one of the key tools in the description of turbulent flows, allowing us to unveil structure out of seemingly chaotic structures in real space. The power spectrum is pivotal for the modelling of turbulence as our understanding of turbulence rests on being able to characterise and link the motions at distinct scales, thereby describing the energy cascade in the inertial range (the length scales in between the scale in which energy is injected and the one in which energy is dissipated). The velocity power spectrum, $P_v(k)$, shows how turbulent kinetic energy cascades from the larger to smaller scales. For shock dominated Burger's turbulence we have that $P(v) \propto k^{-2}$ \citep[e.g.][]{Falceta_2014,2011_Kritsuk}. 
 
The velocity power spectrum is defined in terms of the Fourier transform of the velocity $\hat{\mathbf{v}}$ as: 

\begin{equation}
    P_v(k) = \int\frac{|\hat{\mathbf{v}}|^2}{2} 4\pi k^2 dk
\end{equation}

Thus, integrating $P_v(k)$ over all scales gives us the specific kinetic energy.

\begin{figure}
    \centering
    \includegraphics[width=\columnwidth]{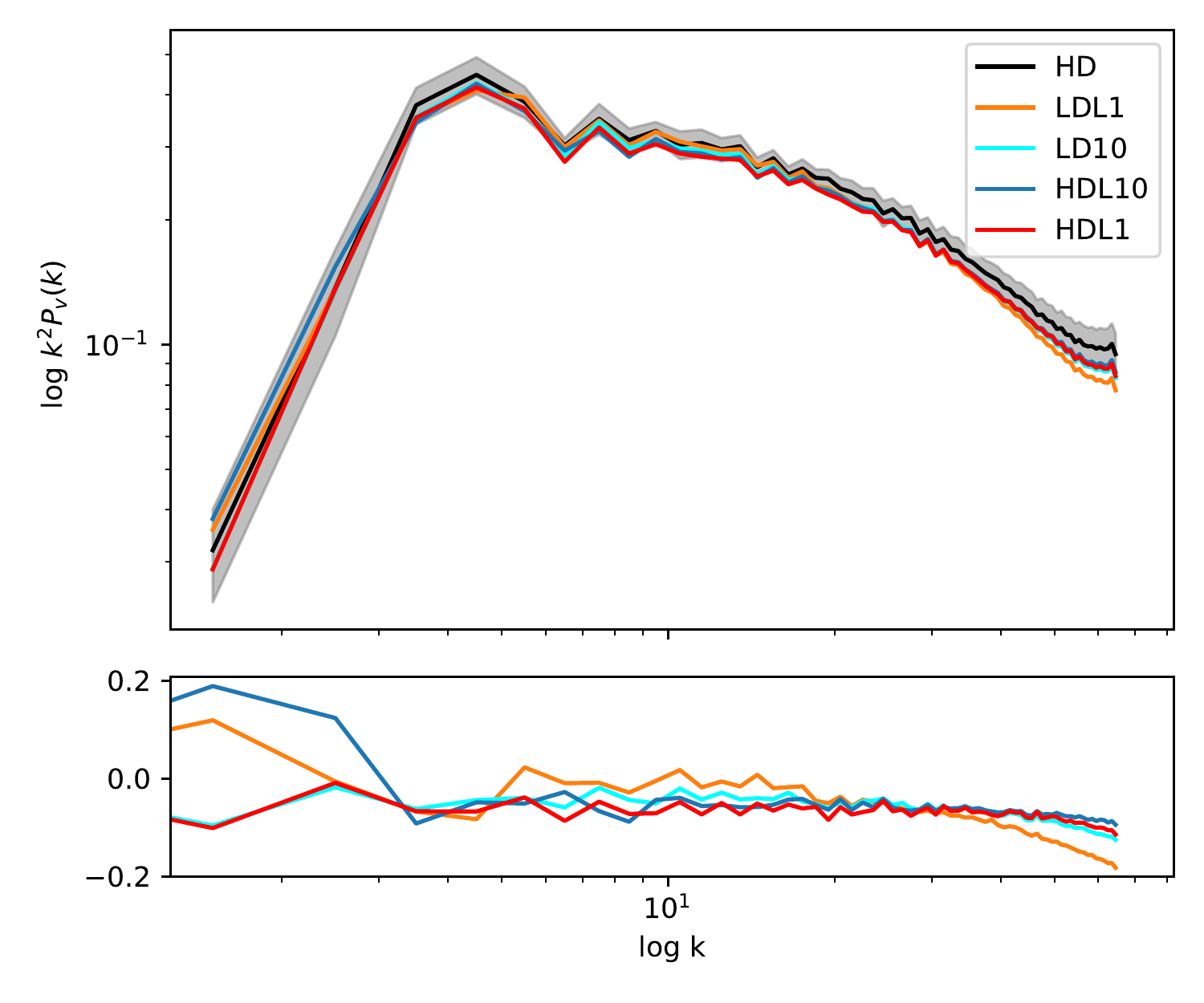}
    \caption{Top: The velocity power spectrum for the RHD runs and the control run. As in Figure \ref{fig: PDF_surf_dens}, the spread for the HD run is again shown by a grey shaded region. Bottom: the fractional change of the average PDF for each RHD run with respect to the average of the control simulation }
    \label{fig:velocity_PS}
\end{figure}

In Figure \ref{fig:velocity_PS} we show the velocity power spectrum compensated by $k^2$. We can see that in none of the runs the power spectrum shows significant distinction from the control run.

\subsection{Without turbulence driving}
\label{subsec: without driving}

\begin{figure}
    \centering
    \includegraphics[width =\columnwidth]{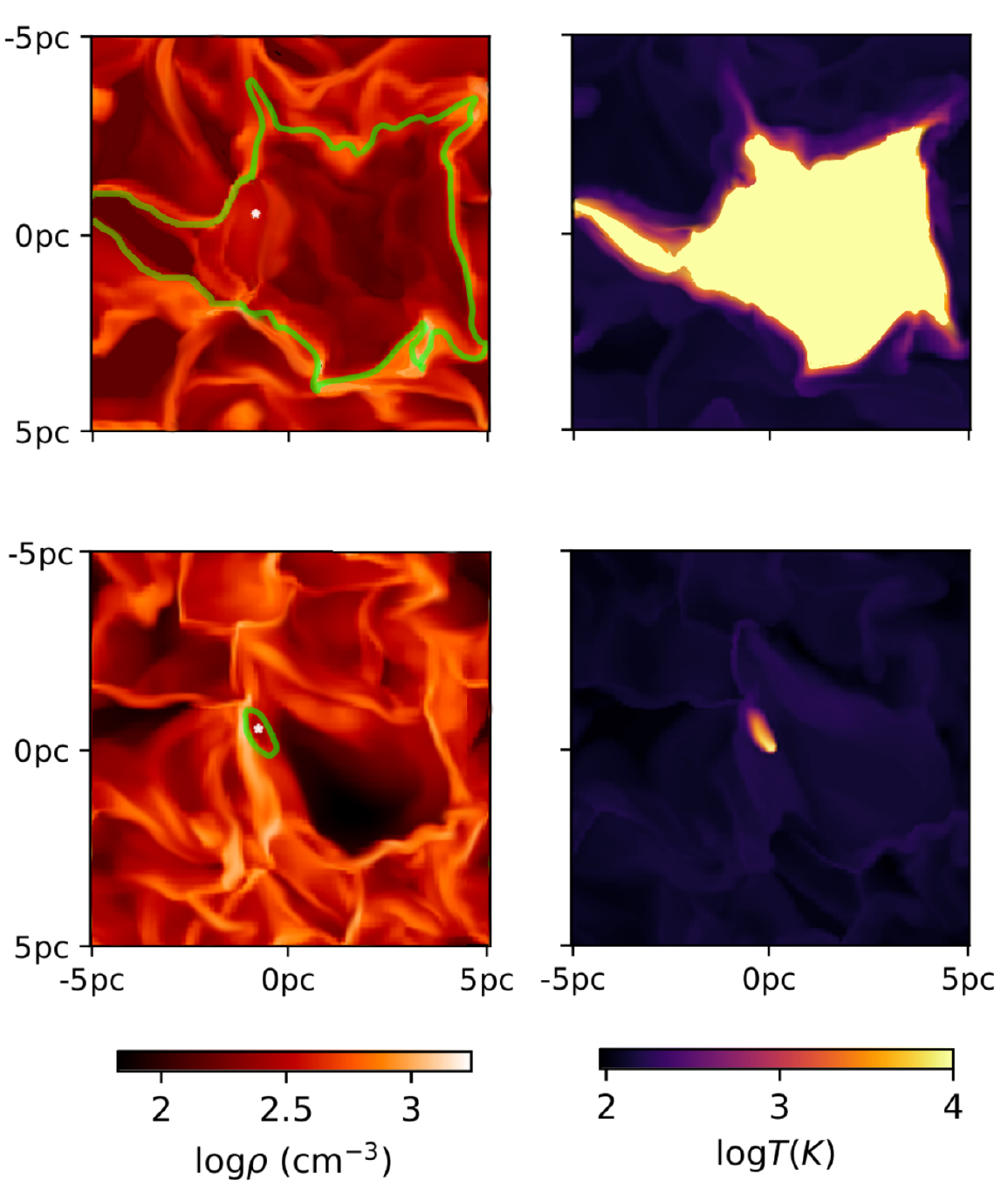}
    \caption{Top: simulation with decaying turbulence. Bottom equivalent simulation for continuously driven turbulence. Both simulations have a source with $Q= 10^{49} \rm{s}^{-1}$ }
    \label{fig:turbulence decay}
\end{figure}

All the simulations presented thus far have a continuous driving of the turbulence during the RHD phase. If the driving is switched off, the filaments start to naturally relax due to the lack of self-gravity. After $\sim 10^6$ yrs, the density contrast within the cloud is significantly diminished. As a result, if an ionizing source is placed within this simulated cloud, the pressure from the hot gas in the h{\sc ii} region alone manages to destroy and push the remaining traces of the filaments more effectively (see Figure \ref{fig:turbulence decay}). This leads to a much larger ionized region than in the case where turbulence is continuously driven. 
This effect was probed by switching off the turbulent driving in our simulation at the same time we introduced an ionizing source. This simulation was equivalent to run HDL1, but after $1$ Myr the H{\sc ii} region formed is much larger than in the run with continuously driven turbulence, as can be seen in Figure \ref{fig:turbulence decay}.

\begin{figure}
    \centering
    \includegraphics[width = \columnwidth]{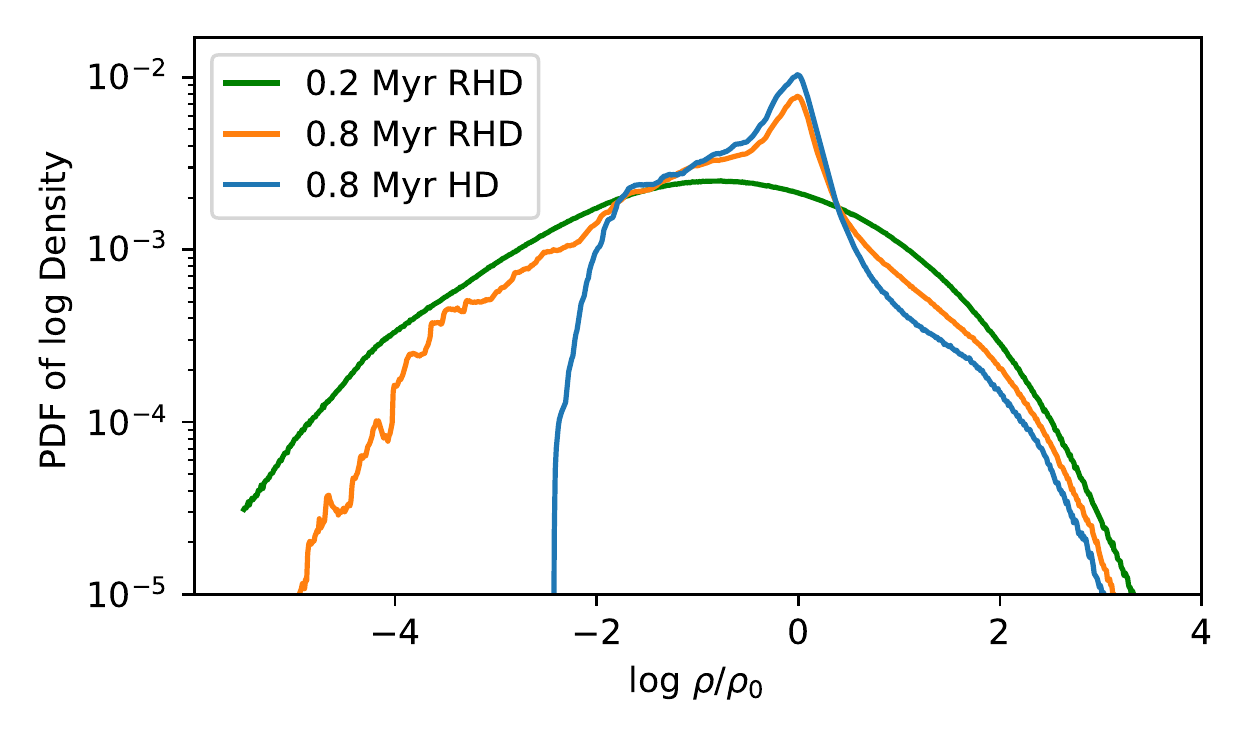}
    \caption{Full PDF (ionized and neutral gas) of decaying turbulence of three snapshots, two of a RHD run and one of a control decaying HD run. Times for the snapshot are counted from the time the source was introduced, which is also the time the turbulence driving is turned off.}
    \label{fig:PDF of decaying turbulence}
\end{figure}

This simulation was only run for 0.8 Myrs from the introduction of the source. This is due to the fact that, although the boundaries of our simulation box are periodic for the hydrodynamics, the boundaries are set to be open for the radiation.
Thus, as soon as our ionizing region starts getting larger than the box size, our simulations are no longer accurate. We plot the PDF for two snapshots of this run (at 0.2 and 0.8 Myrs), as well as the PDF of the snapshot of a control run without radiative feedback in Figure \ref{fig:PDF of decaying turbulence}. It can be seen that, when turbulence driving is disabled, the gas in the HD (control) run starts getting redistributed: the voids begin to be refilled by material and the filaments are no longer as dense as before. This is expected since, without self-gravity to hold dense structures together, gas pressure will eventually restore the system to the uniform density scenario. The same qualitative changes seen in the control run happen for the ionizing run at the same time scale. However, the ionizing pressure from the source manages to keep the voids at relatively low densities, which can clearly be seen by its larger low density wing. Simultaneously, the photoionizing feedback also compresses some gas to higher densities, keeping the high density wing larger than for the HD case. For comparison we also plot the ionization at 0.2 Myrs for the RHD run. At this time we would expect the ionizing source to have already altered the filaments, if it could. The fact that the slices at 0.2 Myrs and 0.8 Myrs are so different indicates that as turbulence decays, the ionizing radiation is able to provide a greater role in the shape of the PDF.

Interestingly, this effect is similar to what we find if turbulence is run with a very low resolution, as then the densest regions in the filament are not properly resolved such that recombination times in the filament are smaller and the filaments are more easily ionized and pushed around as a result of photoionizing feedback.

\section{Discussion}

A number of works within the past few years has tackled the effect of photoionization within molecular clouds and it is interesting to draw a comparison between results. Some of the previous work included only photoionization \citep{Mellema_2006,Dale_2012,Walch_2012, Boneberg_2015}, whereas others have studied photoionization combined with another feedback mechanism, such as radiation pressure \citep{Ali_2018, Kim_2018K} or winds \citep{Dale_2014}, or considered extra physics, like \citet{Geen_2015}.

In many of these works, the turbulent molecular cloud is completely destroyed. However, different studies arrive at very different results, with molecular clouds being sometimes completely dispersed by the ionizing radiation of one B-star \citep{Geen_2015}, one O-star and a few B stars \citep{Ali_2018} or a few clusters of stars \citep{Kim_2018K}, and dispersal time scales for the molecular cloud varying from less than a Myr \citep{Ali_2018} to a few Myrs \citep{Kim_2018K}. 

This work varies from the previous studies in this respect. Firstly, we consider a turbulence-in-a-box type of scenario, similar to \citet{Mellema_2006}, but which significantly differs from most of the recent work \citep{Geen_2015, Ali_2018, Kim_2018K}. Usually, a cloud is simulated by adding a turbulent velocity field to a well defined dense region (often spherical at the start), which represents the molecular cloud and then embedding it in a uniform lower density medium. In our case we do not have a defined cloud edge. Instead, our model is more representative of the environment in the central region of a large molecular cloud.

This choice comes with advantages and disadvantages. In the usual set up, as soon as an ionizing source is introduced, and gas within the molecular cloud heats up, a pressurised bubble is created. This bubble will naturally expand while trying to achieve equilibrium with the cold low density environment where it finds itself in. In that respect, it is not surprising that radiative feedback would lead to the destruction of the molecular cloud unless the self-gravity of the cloud is high enough to withstand the increase in pressure that comes from increasing the inner temperature of the cloud by roughly two orders of magnitude. Often in these simulations, the densest regions are not ionized, and as material is pushed around it by the expanding H{\sc ii} region, neutral fingers, similar to the ones observed in Orion \citep{O'Dell_2020} and other nebulas are formed \citep{Dale_2012, Walch_2012,Geen_2015,Ali_2018}. These are usually seen as indication that the simulation display a somewhat faithful description of molecular clouds.

Neutral fingers can, however, are also seen in turbulence in a box simulation as was shown in the work of \citet{Mellema_2006}, which presented synthetic images of ionized regions formed in turbulence-in-a-box simulations. It should be noted that in \citet{Mellema_2006}, unlike this work, a continuous turbulent driving was not employed and once the source was introduced their boundaries were modified from being periodic to being open allowing gas to escape. Their resultant ionized region quickly grows larger than their simulation box (4 pc$^3$), being thus somewhat similar to the run presented in Section \ref{subsec: without driving}.  In this scenario, what would entail a full dispersal of a molecular cloud is arbitrary. If we define our cloud as being the size of a few parsecs, then in \citet{Mellema_2006} as well as in some scenarios the ionizing feedback would be able to dismantle the molecular cloud. If, however, we consider a cloud being of the order of 100 pc, then our models suggest the H{\sc ii} region would probably remain confined, provided the ionizing source is not located close to the edge of a molecular cloud. In order to fully disperse large regions of molecular clouds, a large number of massive stars would be required.
In reality, molecular clouds are neither completely isolated (with a clear boundary of where they start and end) receiving very little external influence from their surroundings, nor are they part of an infinite turbulent structure, as depicted in this work. The actual impact of ionizing radiation on turbulent molecular clouds can thus only be analyzed when we are able to replicate both the molecular cloud and its surroundings in a single large-scale simulation.

Another important point of comparison between this work and previous works is how we should simulate turbulence within molecular clouds. As shown in Section \ref{subsec: without driving} and as it is the case for the works cited above, turbulence can be allowed to decay over time. In the case of \citet{Boneberg_2015}, they concluded that photoionization from many sources was able to reinstate turbulence within the molecular cloud. In other works, turbulence is also allowed to decay, but often the cloud is entirely dispersed before a turbulent state due to photoionization feedback settles in. In our simulations, the continuous driving of turbulence helps keeping the filaments dense enough to halt the expansion of the ionized region.  It is expected that ionized region are not devoid of turbulence. Supernova are often sufficient to inject enough turbulent energy that can cascade to smaller scales. It is also import to point out that besides supernovae other processes are required to sustain turbulence in molecular clouds to explain observations. A number of simulations have shown that other stellar feedback as winds, proto-stellar outflows and radiation are able to maintain the turbulent state within a molecular cloud \citep{Carroll_2009, Krumholz_2014,Offner_2015,Boneberg_2015}. These processes would maintain some level of turbulence even in ionized regions. Another possible driver of turbulence has been shown to be the interaction of the molecular clouds with the potential of the spiral arms of our galaxy \citep{Falceta_2015}. Contrarily to previous drivers mentioned, the resulting turbulence does not depend on local properties such as star formation rates. If this driving method does take place in nature it would act on ionized and non-ionized gas alike and thus implying that ionized regions should be turbulent.

Lastly, in the our run where turbulence is allowed to decay, the density contrast created by the turbulence driving slowly disappears, and the effect of photoionization is intensified. It is true that in our simulations this effect is also enhanced by the lack of self-gravity, but a similar effect should happen even for simulations with self-gravity (except in the gravitationally dominated densest regions), as can be seen by the control runs (without ionizing sources) in previous works \citep{Boneberg_2015,Geen_2015}. 

The question is, thus, whether turbulence should be allowed to decay or not in simulations trying to analyse photoionization effects in turbulent clouds. To a certain extent this depends on what is the culprit of turbulent driving. If turbulence is driven, at least partially, by local stellar feedback or by other mechanisms that affect ionized and neutral gas alike (as the torques provided by the spiral potential mentioned above) it is possible that turbulence in the clouds is indeed continuously driven and never decays substantially. If it is only driven by more sporadic events, such as a nearby supernova, maybe it can decay over time.

Finally, we point out that, while some change can be seen in the 2D statistics, none of our simulated RHD scenarios vary in a way that could be detected observationally. It is probable that in regions were the photoionization feedback is due to a small number of sources, H{\sc ii} regions will remain largely confined and the statistics of turbulence inferred from observations will not depend on the fact that the interstellar medium is multiphase.

\section{Conclusions}

This study provided an analysis of the effects of photoionization to the statistics of turbulence. We have examined a number of scenarios with sources with different intensities and placed in distinct positions within the turbulent region. The main conclusions can be summarised as follows:

\begin{itemize}
    \item A single radiative source is not able to completely dismantle a high density filament when placed next to it. However, it is possible that multiple massive stars placed at many points within a filament could potentially lead to its destruction.
    \item The effect photoionization has on the gas and on the turbulence statistics depends on where the source is placed, as well as on the source's ionizing flux.
    \item Sources placed in voids create H{\sc ii} regions that occupy a larger volume fraction of the box in comparison to sources placed in high density filaments. However, independently of location, the number of cells that have their density substantially altered is very small.
    \item The effect of the radiative source on the 2D turbulence statistics is minimal, which implies that fairly robust analysis of turbulence can be done even when ignoring the fact that the ISM is multiphase.
    \item If turbulence is allowed to decay, the effects of photoionization are greatly intensified, because the gas pressure makes denser structures more diffuse and, as a result, filaments (and any substructure) are more easily disturbed by the pressure from the warm regions.
\end{itemize}

The results of this work can be greatly improved in the future by considering the effects of magnetic fields and self gravity, both of which are known to play a role in the true distribution of gas in molecular clouds and alter the statistics of turbulence. In particular, self-gravity would allow us to self consistently track the formation of stars and more accurately trace the expected effects of feedback. We plan to address these in future work.

\section{Acknowledgements}
NS acknowledges funding from the Royal Society University Research Fellowship of Anastasia Fialkov and the CAPES funding agency for financial support during her PhD.
BV acknowledges funding from the Belgian Science Policy Office (BELSPO) through the PRODEX project “SPICA-SKIRT: A far-infrared photometry and polarimetry simulation toolbox in preparation of the SPICA mission” (C4000128500). KW and BV acknowledge support by STFC grant ST/M001296/1. Part of this work was  performed  using  the  DiRAC  Data  Intensive  service  at  Leicester,  operated  by  the  University  of  Leicester  IT  Services,  which  forms  part  of  the  STFC DiRAC HPC Facility (www.dirac.ac.uk). The equipment was funded by BEIS capital funding via STFC capital grants ST/K000373/1 and ST/R002363/1 and STFC DiRAC Operations grant ST/R001014/1. DiRAC is part of the National e-Infrastructure.

\section*{Data Availability}

The data underlying this article will be shared on reasonable request to the corresponding author.




\bibliographystyle{mnras}
\bibliography{ref} 








\bsp	
\label{lastpage}
\end{document}